# Totalitarian Technics:
# The Hidden Cost of AI Scribes in Healthcare

Hugh Brosnahan
Bioethics Centre, University of Otago
hugh.brosnahan@postgrad.otago.ac.nz

**Abstract**

*Artificial intelligence (AI) scribes—systems that record and summarise patient–clinician interactions—are promoted as solutions to administrative overload. This paper argues that their significance lies not in efficiency gains but in how they reshape medical attention itself. Offering a conceptual analysis, it situates AI scribes within a broader philosophical lineage concerned with the externalisation of human thought and skill. Drawing on Iain McGilchrist's hemisphere theory and Lewis Mumford's philosophy of technics, the paper examines how technology embodies and amplifies a particular mode of attention. AI scribes, it contends, exemplify the dominance of a left-hemispheric, calculative mindset that privileges the measurable and procedural over the intuitive and relational. As this mode of attention becomes further embedded in medical practice, it risks narrowing the field of care, eroding clinical expertise, and reducing physicians to operators within an increasingly mechanised system.*

## Introduction

Artificial intelligence (AI) scribes—systems designed to record, transcribe, and summarise patient–clinician interactions—are often touted as solutions to the administrative overload faced by healthcare professionals (Tierney et al., 2024, 2025). By automating documentation, these systems, also known as digital or ambient scribes, promise to relieve physicians of clerical tasks and restore their focus to patient care. Yet their integration into clinical settings is more than just a technical innovation; it marks a deeper process in which technology externalises and embeds a particular mode of human attention in the world. Attention is not a passive faculty of selection; it is the act that shapes how the world is disclosed to consciousness. How one attends—narrowly and analytically or broadly and contextually—conditions what can be perceived, valued, and known. Attention therefore carries ethical weight: it determines what we notice and ignore, how we encounter others, and what kind of reality we inhabit. The introduction of AI scribes into a clinical encounter already subordinated to the technological order furthers a process which privileges the measurable, procedural, and representable over the tacit, intuitive, and relational. This paper offers a conceptual analysis, situating AI scribes within a longer philosophical reflection on how technology mediates and reshapes human attention, with implications that are epistemological, ethical, and ultimately metaphysical.

Philosophers have long recognised that the projection of human skill and perception into external forms reflexively shapes both mind and culture. Plato (2002, 274b–277a) warned that writing as *hypomnēsis*—memory externalised—offers only the semblance of wisdom, since what is remembered by inscription is no longer known by experience. Heidegger later redefined this relation in *The Question Concerning Technology* (1977), arguing that modern technology is not merely a collection of tools but a metaphysical mode of revealing (*aletheia*): *Gestell*, or enframing, through which beings disclose themselves as 'standing-reserve'—calculable, controllable, and available. Developing this insight, Gilbert Simondon (1958) conceived of technical artefacts as possessing their own mode of existence. They are not inert instruments but evolving participants in a process of individuation linking human thought, material form, and environment. For Simondon, technics is a phase of becoming in which psychic and collective life exteriorises itself through successive acts of invention—a process he terms *transduction*. Bernard Stiegler (1998), drawing explicitly on Simondon, described this dynamic as *epiphylogenesis*: the externalisation of memory and attention through successive technical supports that shape human temporality and subjectivity—an idea later extended into the digital by Yuk Hui (2016).

These accounts form a continuous reflection on how technology not only serves human purposes but actively configures the conditions of experience, mediating what can be perceived, valued, and known. Yet none of these frameworks fully address the *attentional* dimension of this transformation—the way technology externalises not only skill and memory but the very act of attending itself. Here, the work of Iain McGilchrist (2010, 2021) becomes invaluable. His account of hemispheric asymmetry reveals that our modes of attention condition what kind of world is brought into being. The left hemisphere facilitates the analytic, abstract, and representational; the right, the holistic, contextual, and embodied. Each is essential, yet their balance determines the quality of consciousness and, ultimately, the orientation of a culture. McGilchrist thus provides a bridge between phenomenology and neuroscience, showing that the split within human consciousness mirrors the bifurcation evident in the evolution of *technē*[1]: one mode narrow and controlling, the other integrating and sustaining. His work demonstrates how attentional bias becomes institutionalised in culture and technology, and how attention itself underlies our ways of knowing. Viewed through this lens, AI systems—particularly those automating linguistic and procedural tasks—reflect and amplify a left-hemispheric mode of attention: abstract, procedural, and limited to that which can be represented. Lewis Mumford (1967, 1970) complements this perspective by situating technological development within its moral and civilisational contexts. He distinguishes between *democratic* technics, which enhances human agency and remains compatible with organic scale and diversity, and *totalitarian* technics, which imposes centralised control through standardisation and efficiency. His notion of the 'megamachine'—the fusion of human labour and technical apparatus into a single system of control—offers a prophetic framework for understanding how technological systems reshape institutions and values, including those of modern healthcare.

---

[1] To clarify terms: *technē* refers to embodied, situated know-how—the skilled activity through which a thing is brought forth. Its outputs are technical artefacts, from tools to scripts, which remain answerable to human purposes. The word "technology" itself reflects this evolution: derived from *technē*—embodied skill or craft—and the suffix *-logia*—organised knowledge—it names the systematic coordination and scaling of techniques, rather than the artefacts alone. When artefacts are drawn into this regime, the mode of attention that brought them into being can become inverted, with representation displacing direct encounter with the world.

Together, McGilchrist and Mumford reveal that the technological order is not only a social system but an embodiment of a particular way of attending to the world. When the representational disposition of the left hemisphere becomes operationalised within institutions, it scales into a system of control: the megamachine. The emergence of AI scribes thus belongs to a much older story—the evolving relation between humanity and its instruments. Drawing on these two thinkers, the discussion that follows understands technology as a process that externalises, amplifies, and reflects a particular mode of attention. In medicine, this process promises alleviation while concealing risks beneath its veneer of automated efficiency. Proponents suggest that AI scribes will ease physicians' administrative burdens, enabling greater presence, empathy, and attunement to the patient (GOV.UK, 2025). Yet the very logic that makes these systems efficient—their capacity for representation, classification, and proceduralisation—derives from the left hemisphere's mode of attending. Whether such tools ultimately liberate or constrain therefore depends on how they are integrated into institutional life: whether physicians use the time saved to deepen human contact, or whether efficiency gains merely expand throughput and data obligations. The question is not simply whether AI scribes work, but what kind of world they bring forth within medical practice. For if, as I argue, technology carries an intrinsic tendency toward the totalising logic of the left hemisphere, careful discernment is required to ensure that such instruments do not, in relieving doctors of clerical tasks, also relieve them of the conditions necessary for genuine care. As such, this analysis is concerned with how technology mediates our relation to Being itself—how the world appears to us depending on how we attend. As medical praxis bends further toward the left hemisphere's mode of attending, physicians risk being reduced from practitioners of an art to mere operators within an increasingly mechanical system—ghosts in the machine.

## From Minds to Machines

As outlined in the introduction, this paper situates its analysis within a philosophical lineage that, broadly stated, conceives technology as the externalisation of mind. Stiegler (1998, p. 140) captures this notion with the term *epiphylogenesis* (*epi*, 'upon,' + *phylē*, 'species,' + *genesis*, 'becoming'): the process by which experience is retained and transmitted through technical artefacts—tools, writing, images, and now digital code. These artefacts preserve memory outside the body, forming a new evolutionary archive. Humanity thereby inherits not only genes but technical memory: the accumulated knowledge and experience of prior lives. For present purposes, I propose to develop Stiegler's insight. Epiphylogenesis concerns the exteriorisation of memory and skill; however, when *technē* is formalised into repeatable methods and scaled through institutional systems, we witness the externalisation not merely of experience but of a particular mode of attention. I use the term *epidianoesis* (*epi*, 'upon,' + *dianoesis*, 'discursive reasoning') to describe this projection of the analytic intellect: instrumental, fixed, and representational. This emphasis draws on McGilchrist's view that attention is not one cognitive function among others but the precondition of experience—the act through which the world is disclosed to consciousness. In this light, epidianoesis marks the point where a left-hemispheric mode of attending becomes inscribed into technics themselves, such that the world increasingly appears according to its own manner of disclosure.

As McGilchrist observes (2010, p. 461), "the divided nature of our reality has been a consistent observation since humanity has been sufficiently self-conscious to reflect on it." His account stands in continuity with an insight traceable throughout the history of philosophy: Plato's distinction between *noēsis* and *dianoia* (Republic 511d–e) contrasts intuitive apprehension with discursive reasoning; Heidegger's analysis of *Zuhandenheit* and *Vorhandenheit* (1962) distinguishes our engaged, practical encounter with the world from the detached, objectifying stance of calculation; and Arendt's recovery of the Kantian distinction between *Vernunft* and *Verstand* (1978) restores the difference between reason as the search for meaning and intellect as the pursuit of explanation. Each identifies a polarity within the mind: one receptive, holistic, and oriented toward understanding; the other analytic, representational, and oriented toward control. McGilchrist gives this perennial intuition an empirical and neurophenomenological foundation. His central thesis, grounded in research on hemispheric lateralisation, holds that the two hemispheres mediate complementary yet divergent modes of attention—a division rooted less in *what* they do than in *how* they do it. This bifurcation is thought to have evolved as a solution to a primordial challenge faced by all sentient life: the need to sustain both narrow focus for grasping and manipulation and broad awareness for context and survival. While rudimentary traces of this division exist across species, it reaches its culmination in humans, whose bipartite brain enables two distinct orientations to reality and, consequently, two divergent ways of encountering and bringing the world into being for us (Ocklenburg & Güntürkün, 2012).

Before proceeding, two further thinkers warrant mentioning, both for context and for contrast. The first is psychologist Julian Jaynes (2000), whose *The Origin of Consciousness in the Breakdown of the Bicameral Mind* proposes that introspective consciousness first arose in Homeric Greece. When Homeric heroes 'heard' divine voices, he argues, they were perceiving their own intuitive thoughts—generated by the right hemisphere—as external commands. Drawing parallels with auditory hallucinations in schizophrenia, Jaynes holds that consciousness emerged from the collapse of a previously 'bicameral' mind, as the analytic left hemisphere gained sudden access to the intuitive right. McGilchrist (2010, p. 260) credits Jaynes with a profound insight—the link between divine voices, changing consciousness, and hemispheric function—but reverses the causal order. For Jaynes, consciousness arose through the merging of the hemispheres; for McGilchrist, it arose through their separation. The 'voices of the gods' were not the result of new integration but of a loss of it: intuitions once seamlessly acted upon became alien and objectified as external. What had been transparent now appeared as 'other.' This detachment made reflection, empathy, and self-awareness possible. Where Jaynes saw consciousness as the product of fusion, McGilchrist sees it as the gift of a more pronounced division.

A second, more contemporary interlocutor is Nobel laureate Daniel Kahneman, whose *Thinking, Fast and Slow* (2011) also posits a dual architecture of thought. Yet, as McGilchrist (2021, p. 739) notes, Kahneman's 'System 1' and 'System 2' divide the brain *horizontally*—fast versus slow processing—rather than *vertically* along hemispheric lines. In McGilchrist's work, we find that both hemispheres engage in heuristics and deliberation, but with distinct tendencies: the left favours rapid, target-driven judgment and clings to its conclusions, while the right remains cautious, integrative, and self-correcting. Under time pressure, the right prefrontal cortex's balancing role weakens, allowing the left hemisphere's quick, biased heuristics to dominate. What Kahneman describes as 'fast thinking,' then, corresponds less to speed *per se* than to left-hemispheric overreach—a mode of cognition marked by premature closure and resistance to revision. Given the importance assigned to attention in this paper, the manner in which the brain is divided matters. As McGilchrist (2021, p. 35) observes, "there are emergent phenomena from the whole interconnected system at the hemisphere level, and neuroscience increasingly recognises that we should think in terms of complex widespread networks." The hemispheres exemplify such networks *par excellence*, being "maximally connected within themselves" and *each alone capable of sustaining consciousness*. Moreover, while it is important to emphasise that references to the left and right hemispheres correspond to actual brain structures, they also serve as synecdoches for broader cognitive tendencies within our unified experience of consciousness. As philosopher John Cottingham (2019: 363) points out, "McGilchrist makes it clear that the relevant distinction should be construed, as it were, as an adverbial one rather than a functional one." The key differences are not found in their particular 'functions'—important as they are—but rather in the *manner* in which the hemispheres engage with the world. Framing cognitive division at this level—rather than in abstract processing modes—offers a more coherent and biologically grounded understanding of how distinct attentional styles shape our engagement with the world.

According to hemisphere theory, the left hemisphere's mode of attention *constructs* a world of "isolated, fragmentary elements" that are "easily manipulated," detached, and abstract—an "inanimate universe and a bureaucrat's dream" (McGilchrist, 2021, p. 31). This attentional style is confident yet shallow: effective for short-term manipulation but blind to context, complexity, and meaning. The world it discloses—a realm reduced to static, divisible, and controllable parts—closely parallels the order envisioned by Mumford's *megamachine*: a total system in which human activity is absorbed into rigid, technocratic structures oriented toward control and optimisation. Both emerge from the same underlying disposition—the will to mastery that subordinates organic life to mechanical regularity, and meaning to function. As we shall see, the relationship between what Mumford calls *totalitarian technics* and left-hemispheric attention is critical for understanding the hidden risks of our technological society, particularly in medicine. Conversely, Mumford's notion of *democratic technics* finds its analogue in the right hemisphere's orientation, which opens us to a world "almost infinitely more complex" (McGilchrist, 2021, p. 31)—a world fluid, relational, and alive, in which "nothing is ever static or detached from our awareness," and wholes exceed the sum of their parts. This mode of attention, though less efficient and difficult to formalise, remains more faithful to reality itself.

Given these initial accounts of the worlds brought into being by the hemispheres, one might propose rejecting the left hemisphere's vision outright. Yet this very impulse reflects the left hemisphere's 'either/or' approach, which favours simplistic and unambiguous conclusions. The right hemisphere, in contrast, embraces a 'both/and' perspective, understanding, as generalists do, the need for the specialist—if concurrently appreciating the specialist's limitations. While specialisation is an indispensable intellectual tool, indispensable must not be conflated with all-encompassing. The analytic faculty is, in this sense, both necessary and useful for the business of getting about in the world, provided its 'take' is reintegrated by the *Gestalt*-oriented faculties of the right hemisphere.

This dynamic points to an inherent priority of mental activity—one that follows a progression from right to left, then, crucially, back to the right again. At the neurophysiological level, it is the right hemisphere that first attends to the unfamiliar and unmapped aspects of perception. "Anything newly entering our experiential world instantly triggers a release of noradrenaline—mainly in the right hemisphere" (McGilchrist, 2010, 40). Indeed, anything novel, from experience to learning new information or developing unfamiliar skills "also engages right-hemisphere attention more than left, even if the information is verbal in nature" (McGilchrist, 2010, 40). However, once whatever it is that the right hemisphere has been attending to becomes familiar, the left hemisphere takes over, even in domains that are predominantly facilitated by the right hemisphere, such as music. The left hemisphere's expertise lies in its capacity to *re*-cognise what comes from the right hemisphere with its narrow beam attentional capacity and its facility for representation, categorisation, and manipulation. Its abilities in this domain allow it to render the previously unknown 'known,' in an abstract manner, before submitting its necessarily reductive conclusions back to the right hemisphere—the house of global attention—for reintegration.[2]

---

[2] The relationship between the hemisphere's calls to mind Bertrand Russell's (1945, 83) observation that "a stupid man's report of what a clever man says can never be accurate, because he unconsciously

This process of reintegration is vital and highlights a key aspect of McGilchrist's theory: hemispheric asymmetry is not merely physiological but epistemological. The hemispheres do not differ only in the kinds of information they handle, but in the *manner* in which they reveal and make sense of reality. Crucially, however, the left hemisphere's representational stance remains dependent on the right hemisphere's broader, context-sensitive mode of disclosure:

> What the left hemisphere sees as finished, perfect, single, abstract, detached, motionless, beyond space and time, is virtual only; a reduced re-presentation at an instant outside time of what according to the right hemisphere is always evolving, ever both self-differentiating and self-unifying, and involved with the process of creation and what it creates (McGilchrist, 2021, 1235).

To borrow a familiar metaphor, if the left hemisphere provides the map, the right hemisphere surveys the terrain.[3] Without adequate awareness of the territory, the fidelity of the map is bound to suffer.

Epistemological asymmetry is not merely a theoretical concern; it manifests in concrete ways when the relationship between the hemispheres is disrupted due to pathology. When individuals suffer damage to one hemisphere, their mental approach shifts accordingly, providing a stark illustration of this disruption in action. Those "with a right-hemisphere lesion (therefore relying on their intact left hemisphere) tend to start with the pieces and put them together to get the overall picture, whereas those with a left-hemisphere lesion (relying on their right hemisphere) prefer a global approach" (McGilchrist, 2010, 28). Without the contextualising faculties of the right hemisphere to provide the 'big picture' and common sense, the left hemisphere's fragmented and narrowly focused approach to theory and 'map-making' leaves affected individuals vulnerable to 'losing the plot'—literally. This phenomenon is often manifested by individuals on the schizoid-autistic spectrum. In his neuroimaging research, McGilchrist observes that the typical asymmetric pattern of brain activity is either absent or reversed in cases of schizophrenia. These affected areas "are important in language, social functioning and the business of making sense of the world at the highest level—areas of life that just happen to be those in which subjects with schizophrenia encounter major problems" (2010, 26). Elsewhere, he writes that "the confluence of phenomenology between right hemisphere-damaged and schizophrenic patients, both how they perceive the world and come across to others, is extraordinary" (2010, 309). Research supports this observation, showing that schizophrenic patients are "statistically identical to patients with right brain damage and robustly different from those with left brain damage" when compared on a test of social and emotional interpretation and expression (Ross et al, 2001). Furthermore,

> both groups tend to neglect context, and therefore have difficulty appreciating the 'discourse elements' of communication. Both find narrative difficult to understand, and difficult to generate. Both complain of being unable to read a book or watch a film, because they cannot integrate the storyline over time… They have similar problems dealing with the implicit—not just tone of voice,

---

translates what he hears into something he can understand." To do its job, the left hemisphere must abstract and categorise what the right hemisphere *presences*, thereby risking a similar distortion—reducing the richness of experience to a framework it can control, thereby diminishing the original insight.

[3] Philosopher Alfred Korzybski (1931), who coined the phrase "the map is not the territory," emphasises that representations of reality (maps, words, models) are not the same as reality itself. They are abstractions or symbols that can guide understanding but are inherently limited and incomplete. In a similar vein, the statistician George Box observes that "all models are wrong, but some are useful" (Box and Draper 1987, 424).

but tone of any kind, including humour, irony or sarcasm; facial expressions and body language; and other implicit processes such as understanding metaphor (McGilchrist, 2021, 309).

This neglect of context underscores the critical role the right hemisphere plays in providing the broader, integrative perspective necessary for navigating social and narrative complexities, such as the doctor-patient relationship.

Also significant is what McGilchrist (2021, 176) describes as "an excess of confidence and a lack of insight"—a hallmark of left hemisphere dominance. This tendency is closely linked to the left hemisphere's propensity for confabulation and its susceptibility to delusion, which occurs at a rate of 3:1 compared to the right hemisphere (2021, 178). These confabulating tendencies stem, in part, from its aversion to ambiguity and its insistence on consistency, crucial features for a mental faculty that has evolved to facilitate our *grasping* things—in all senses of the word. As one study concludes, the left hemisphere "appears to detest uncertainty; it creates explanations and fills in gaps of information in order to build a cohesive story and extinguish doubt" (McGilchrist, 2021, 154). Neuropsychologist Michael Gazzaniga similarly observes that when the left hemisphere lacks knowledge, it refuses to admit ignorance. Instead, it "would guess, prevaricate, rationalize, and look for a cause and effect, but it would always come up with an answer that fit the circumstances" (McGilchrist, 2021, 91). This disdain for uncertainty drives the left hemisphere to jump prematurely to conclusions, which can prove a dangerous tendency if the right hemisphere's contextualising, 'devil's advocate' approach fails to intervene. "When the right hemisphere is damaged, patients' hypotheses become rigid and impermeable to conflicting evidence … These findings suggest that a functioning right hemisphere is necessary to detect inconsistencies and update flawed hypotheses accordingly" (McGilchrist, 2021, 173).[4] It is unsurprising then that, as McGilchrist (2021, 180) writes, "virtually all delusional syndromes are more commonly the result of right hemisphere than left hemisphere dysfunction … and no delusional syndrome is commoner in left than right hemisphere dysfunction"

**Epidianoesis**

These intrinsic asymmetries take on new ethical significance once we recognise that emerging clinical technologies increasingly reflect—and amplify—the left hemisphere's mode of attention. The implications of these findings for medical practice (and indeed for the development of AI) are striking. The pathologies McGilchrist describes—overconfidence, lack of insight, confabulation, and delusional coherence—find an uncanny parallel in the limitations of AI systems, such as large language models (LLMs). In technical terms, these manifest as 'hallucinations,' 'overfitting,' and 'premature closure'—different names for the same underlying pattern: the production of plausible yet unfounded representations. This resemblance is not merely metaphorical but structural. Both AI systems and the left hemisphere share a common mode of operation shaped by the latter's attentional disposition: privileging coherence over truth, closure over openness, representation over reality. I draw attention to these correspondences not to anthropomorphise AI, but because these systems now concretise the left hemisphere's logic at scale. Insofar as this externalisation reproduces the left hemisphere's cognitive architecture, it inherits not only its strengths—abstraction, systematisation, efficiency—but also its characteristic weaknesses: rigidity, confabulation, and delusional certainty. Absent the right hemisphere's grounding and corrective capacities—the ability to test the map against the terrain—technological systems remain ontologically confined to the left-hemispheric world of representations alone.

---

[4] Lest we dismiss the relevance of these findings for people unaffected by clinical pathologies, note that, according to McGilchrist (2021, 154), "subclinical delusional beliefs [are] directly tied to left hemisphere activation in *normal participants*—not just psychiatric patients."

Having established the neurocognitive foundations of attentional asymmetry, we can now trace how these same dynamics scale outward into the technological and institutional structures of civilisation—a trajectory Lewis Mumford (1967) charts in his account of the megamachine. In *The Myth of the Machine*, he describes this phenomenon as more than a collection of mechanical devices; it encompasses the vast organisation of society itself—composed of flesh-and-blood components—structured around principles of efficiency, standardisation, and control, all enabled and reinforced by technology. According to Mumford, the megamachine first appears in embryonic form at the end of the fourth millennium BC, with the construction of ziggurats and pyramids, before assuming more recognisably modern contours in the medieval period, when inventions such as watermills, windmills, and mechanical clocks introduced a mechanistic approach to harnessing natural forces. Again, this account is more than just a metaphor:

> If a machine be defined … as a combination of resistant parts, each specialized in function, operating under human control, to utilize energy and to perform work, then the great labor machine was in every aspect a genuine machine: all the more because its components, though made of human bone, nerve, and muscle, were reduced to their bare mechanical elements and rigidly standardized for the performance of their limited tasks (Mumford, 1967, p. 191).

A contemporary of Mumford's, the philosopher of technology Jacques Ellul (1964), reaches similar conclusions in *The Technological Society*. For Ellul, *technique* denotes not mere tools or machines but a self-perpetuating system of means detached from ends: a dynamic in which every technical possibility demands realisation simply because it is possible. Against the view of technology as a neutral set of instruments shaped by human needs and values, Ellul depicts it as an autonomous force that conditions human life and perception in its own image. Humanity becomes shaped by technique rather than the other way around. Unlike Mumford, however, Ellul denies the possibility of democratic technics. In McGilchristian terms, he describes a world already 'colonised' by the left hemisphere's vision, where the technical impulse becomes ontologically self-justifying—unable to imagine re-integration within the relational, value-laden vision of the right.

Ellul's analysis captures the pathological culmination of a process that Mumford still believed could be redirected. Where Ellul sees only the self-reinforcing autonomy of technique, Mumford discerns within technics a latent duality—a tension between what dehumanises and what sustains life. From its inception, technics has contained two opposed tendencies: one coercive, hierarchical, and dehumanising; the other life-enhancing, creative, and oriented toward human flourishing. The megamachine represents the triumph of the former—the realisation of technics as a system of control. As outlined above, Mumford (1967) names these opposing forms *totalitarian* and *democratic technics*: the former centralises power, fragments knowledge, and subordinates human life to industrial and economic imperatives; the latter seeks to integrate technical development with moral value, human plurality, and ecological balance. Democracy, he writes, "favours the whole rather than the part" and "can be embodied only in living human beings who interact freely as equals within small, face-to-face communities"—the antithesis of "anonymous, depersonalised forms of mass organisation" (1967, p. 236).

Modern healthcare, for all its technical brilliance, exemplifies this tension. Few would wish to return to medicine before the advent of antibiotics, yet by Mumford's criteria, contemporary healthcare increasingly conforms to the totalitarian model: technically sophisticated yet dehumanising through its bureaucratic sprawl, hyper-specialisation, and the fetishisation of efficiency. The inversion of values is evident in the disproportionate growth of administrative roles. Between 1975 and 2010, for example, the number of healthcare administrators in the United States expanded by more than 3,000 per cent—thirty-two times faster than the physician workforce, which increased by only 150 per cent (Pentherby & Pentherby, 2020). Such tendencies, Mumford contends, are rooted in an intellectually scientific lineage: the mechanistic worldview inherited from Galileo and Kepler and fully realised in Descartes' conception of a world governed by abstract rationality.

> The missing elements in Descartes' grossly over-simplified mechanical model, and in the scientific outlook that … has taken that model over, are history, symbolic culture, mind—in other words, the totality of human experience not simply as known but as lived … To heed only the abstractions of the intelligence or the operations of machines, and to ignore feelings, emotions, intuitions, fantasies, ideas, is to substitute bleached skeletons, manipulated by wires, for the living organism. The cult of anti-life secretly begins at this point, with its readiness to extirpate organisms and contract human wants and desires in order to conform to the machine (Mumford, 1970, p. 91).

This paradigm, foundational to the modern megamachine, privileges control, quantification, and efficiency over the human scale of experience. It leaves no room for that which cannot be abstracted or reduced. Insofar as we live under its aegis, the challenge is to re-imagine medicine—and society more broadly—in a more democratic form: one that balances technological progress with the holistic vision of the right hemisphere, grounded in human relationships, proportionality, and values.

Mumford's megamachine is not only a historical phenomenon but the manifestation of a deeper metaphysical condition: the ascendancy of a particular mode of attention. Its totalising logic is the external counterpart of an inner division in human consciousness—the triumph of the grasping, representational orientation over the participatory and responsive. Here McGilchrist's framework proves crucial. Insofar as technology constitutes externalised attention, it embodies the attention of the left hemisphere. To describe technology as left-hemispheric in essence is not to reduce it to a neurological process but to identify the mode of world-disclosure it instantiates. If technology is understood as the externalisation of that which can be represented—as the projection of models, symbols, and procedures into material form—then it enacts, by its very structure, the left hemisphere's mode of disclosure. From the earliest accounting tablets and bureaucratic scripts to contemporary machine-learning systems, technology fixes what is ephemeral and renders what is implicit explicit. In this sense, it does not merely express a left-hemispheric attitude; it is that attitude made concrete, manifesting the drive to capture, predict, and control—the same drive that, when unchecked, mistakes its representations for the whole of reality. This orientation reveals something essential about the internal dynamic of technology itself, leading directly to Mumford's distinction between democratic and totalitarian technics. Because technology externalises left-hemispheric attention, its innate tendency is toward the totalitarian; it harbours, ontologically, a totalising drive—it reveals Being under the aspect of control. To render our use of technical artefacts genuinely democratic, we must break the spell of representational mastery that informs the very ontology of our instruments. Doing so requires a revolutionary stance of attention—a *metanoia* in the truest sense: a turning of mind from control to participation, from representation to reality. Yet such continual conversion is arduous even for the critically attentive mind; and within a civilisation whose culture and institutions increasingly reflect the disposition and values of the megamachine, it becomes a truly Daedalian task. We find ourselves navigating a labyrinth of our own devising, its walls rising ever higher with each new act of technological ingenuity.

## Theory in Practice

The Daedalian predicament is perhaps nowhere more apparent than in contemporary medicine. AI scribes are promoted as remedies for the bureaucratic overload that increasingly constrains clinical practice, promising to restore physicians' time and attention to their patients (Heidi Health, 2025). Yet, as argued in the preceding discussion, technological 'solutions' can never extend the full spectrum of human capacities, for they are bound by the logic of the mode from which they arise. Because technology externalises the attention of the left hemisphere (epidianoesis), it cannot escape its *Weltanschauung*, whose ontology is one of representation, categorisation, and control. From within the all-encompassing architecture of the megamachine, however, this ongoing transformation is difficult to perceive. To be sure, it is possible to imagine AI scribes being integrated in a manner consonant with Mumford's democratic technics: tools that augment rather than replace human capacities. In such a scenario, by automating the drudgery of documentation, physicians might indeed reclaim the time and attentional space required for genuine presence with their patients. This is the promise—and it is not without merit. However, the difficulty lies in sustaining such democratic use within an institutional environment already ordered by the totalitarian values of the megamachine. The very features that make AI scribes attractive to physicians and administrators—their capacity for quantification, proceduralisation, and optimisation—also make them vectors of the drive towards standardisation and efficiency. Without vigilance, the time they appear to free will be quickly reabsorbed into the bureaucratic logic that produced the problem in the first place.

Tierney et al.'s (2025) follow-up report, *Ambient Artificial Intelligence Scribes: Learnings after 1 Year and over 2.5 Million Uses*, illustrates the point. The authors note that the majority of physicians (84 per cent)—and patients (56 per cent)—reported positive experiences with AI scribes. As technical artefacts integrated into the existing technological system, they demonstrably reduced physician workload producing aggregate time savings of more than 15,700 hours—equivalent to 1,794 working days—over a single year compared with non-users. Notably, almost 40 per cent of patients reported that physicians spent more time than usual talking to them, while nearly 50 per cent felt that physicians spent less time looking at their computer screens. Though these last figures hardly constitute a revolution, they nevertheless signal the latent democratic potential of AI scribes: the possibility that automation may create the attentional space required for genuine clinical presence. The question is whether this potential can be sustained against the gravitational pull of the broader evaluative regime—the framework the encompassing technological order already presupposes. Aside from these potentially meaningful improvements in the physician–patient relationship, the metrics by which progress is declared—time saved, notes generated, reduced cognitive load—are themselves artefacts of the left-hemispheric paradigm. Success is assessed according to the values the megamachine privileges: efficiency, throughput, and the quantification of experience.

From the perspective of hemisphere theory, this is precisely the danger. The institutionalisation of AI scribes risks further entrenching a left-hemispheric style of attention under the guise of progress. The promise is self-defeating because it seeks liberation from bureaucratic constraint by reinforcing the very logic that produced it: efficiency, standardisation, and control. This dynamic—our growing reliance on technological interventions to optimise efficiency—discloses the deeper logic of epidianoesis: the outward drift of the left hemisphere's attention, solidified into the technical and bureaucratic architectures of the megamachine. Indeed, at this stage, the process of externalisation reaches a reflexive turn. What began as epidianoesis—the projection of the left hemisphere's attention into the world—culminates in *metadianoesis*: the self-enclosure of the mind within a world of its own representations. The prefix *meta* signifies both 'going beyond' and its subsequent reflexivity—thought turning back upon itself having populated the external world with its constructions. This is what McGilchrist (2024) means when he warns that "the left hemisphere's vision of a lifeless, mechanical, two-dimensional construct has been externalised around us to such a degree that when the right hemisphere checks back with experience, it finds that the left hemisphere has already colonised reality." Metadianoesis, then, names the final stage of this trajectory—the point at which technology no longer merely expresses left-hemispheric attention but begins to reconstitute the world in its image.[5]

As mentioned earlier, this process is particularly visible in modern healthcare, increasingly defined by bureaucratic expansion and economic imperatives. A 2020 survey found that European physicians spent fully half their working time on administrative tasks (Stewart, 2022). That same year, paediatric neurologist Pedro Weisleder (2023, p. 262) reported that "38.2% of US physicians had at least one manifestation of burnout," a figure that soared to 68.2% the following year. In his commentary, Weisleder notes a growing consensus among clinicians that echoes the concerns developed here: administrators impose rigid, profit-driven rules *detached* from medical realities, restrict clinical autonomy, disregard the clerical burden, and expand institutional bureaucracy without regard for staffing constraints. The linguistic texture of healthcare mirrors this shift. As Rosemary Rizq (2013, p. 21) observes, its vocabulary now resembles Orwell's *Newspeak*: a dialect of efficiency in which phrases such as "competence frameworks," "evidence-based interventions," "risk assessment," and "outcome-led services" displace the organic language of human care. Given the gargantuan dimensions of the megamachine, it is not surprising to find the same linguistic turn pervading multiple domains, including scientific and academic discourse. Sociologist Donald Hayes (1992) identifies this "drift towards inaccessibility" as beginning in the late twentieth century, corresponding with the rapid expansion of the technological system organised around computation. He laments that the "erection of higher and higher barriers to the comprehension of scientific affairs must surely diminish science itself." The rising adoption of large language models is unlikely to mitigate this trend. As their influence grows (Gray, 2024), scientific papers will increasingly address not intelligent human readers but the machines that parse them; data replace description, and life itself is translated into the syntax of analysis. Likewise, work by Kousha and Thelwall (2025) on how LLMs are changing the language of academic papers suggest how deeply metadianoetic logic—representation feeding upon representation—has taken hold.

This reflexive enclosure is already operational in contemporary practice. The adoption of AI scribes should therefore not be viewed as an isolated innovation but as the latest episode in a longer historical trajectory: the transformation of healthcare into an increasingly technical form.[6] The result is a world apprehended as system, resource, and standing-reserve. From the written case note to the electronic record, AI scribes extend the same gesture, translating the complex interpersonal temporality of the doctor–patient encounter into codified data streams. In so doing, they reveal the totalitarian tendency of technics described by Mumford—the drive to subordinate living processes to the values of efficiency, standardisation, and control. Such systems do more than merely assist the physician; they reorganise and delimit the field of care according to what can be represented. The ethical question, then, is not whether AI performs this task competently, but what kind of world—and what kind of medicine—it brings forth. What is diminished in this process, because it cannot be automated or enhanced along technical lines, is the right hemisphere's capacity for integrative, intuitive, and embodied attention. This diminishment is especially consequential in the clinical encounter, where genuine understanding depends not only on the literal content of speech but on tone, gesture, silence, and the shared relational field through which emotional and contextual meaning arises. These are dimensions of care that no system of representation can capture. For, while the left hemisphere excels at syntax, categorisation, and abstraction—the "technical machinery" of language, it is the right hemisphere that mediates the higher linguistic functions: understanding tone, irony, metaphor, and emotional significance (McGilchrist, 2010, p. 70).

---

[5] Stiegler warns that when the technical dimension of experience goes unexamined, "the outcome is the pauperisation, the impoverishment, the starvation of consciousness," a "disastrous de-spiriting" born of "inattention to questions of objective memory" (2011, p. 86). This marks the point at which epiphylogenesis becomes self-enclosing—what I term metadianoesis: the reflexive exhaustion of attention within its own technical retentions.

[6] Historian of medicine Jeremy A. Greene traces this trajectory in detail, showing how each new medical medium—from the telephone and early telemedicine to digital diagnostics—has promised liberation from administrative or cognitive burdens while deepening medicine's dependence on technological mediation (*The Doctor Who Wasn't There: Technology's Promise and Peril in the*

Anthropologist Gregory Bateson (1973, p. 340) terms this domain the paralinguistic: "signals, such as bodily movements, involuntary tensions of voluntary muscles, changes of facial expression, hesitations, shifts in tempo of speech or movement, overtones of the voice, and irregularities of expression." Understanding a patient's concerns demands the ability to perceive the subtle nuances not only of what is said but also how it is said—including what remains unspoken. It is within this relational field that empathy, trust, and diagnostic insight emerge. When attention is displaced into the technical apparatuses that support AI transcription, these paralinguistic dimensions are diminished or missed entirely. The physician becomes less an embodied interlocutor and more an operator within an informational system—an "Organization Man," in Mumford's (1964, p. 279) terms, whose existence is reduced to that of a "depersonalized servo-mechanism," reflecting the megamachine's totalising logic. What must be rendered democratic, therefore, is not technology itself but the conditions under which technical artefacts are employed. No correction internal to the technological order can suffice, for its very ontology—the drive toward control, optimisation, and representation—already reflects the left hemisphere's stance. Only through the right hemisphere's broader mode of attention can technical artefacts be brought back under the primacy of *technē*, reintegrated into a genuinely humane practice.

There is a further danger here, already attested to and documented within the literature: the risk of automation-induced cognitive atrophy. Research on learning and cognition illustrates this danger. Studies comparing handwriting and typing show that manual writing activates neural 'circuits' associated with comprehension, memory, and conceptual integration (Rosen, 2024). Students who write notes by hand perform better on conceptual questions than those who type verbatim transcripts (Mueller & Oppenheimer, 2014). These findings affirm McGilchrist's claim that embodied experience grounds higher forms of understanding. As handwriting gives way to typing, and typing to dictation, layers of embodied engagement are stripped away. AI scribes represent the next stage in this abstraction: a further disembodiment of practice. It is unsurprising, then, that Macnamara et al. (2024) find that while AI may enhance short-term performance, it undermines users' ability to monitor or sustain their own skills. Vasconcelos et al. (2023) similarly report that people "often accept incorrect AI outputs without verification," a form of automation-

---

*Age of Digital Medicine*, 2022). His account underscores that AI scribes represent continuity rather than rupture in this long history of the technical reconstitution of care.

induced complacency. The very features that make AI scribes efficient—speed, fluency, reliability—invite overreliance, particularly in overstressed clinical environments where physicians lack time to cross-check AI-generated summaries. These effects are not limited to cognitive skill but extend to the moral and relational fabric of practice. Lee et al. (2019) found that the presence of screens during interpersonal exchanges correlates with higher rates of anxiety and depression, lower empathy, and diminished openness and conscientiousness. The same devices that increase administrative efficiency simultaneously erode the attentional conditions of genuine encounter. In medicine, this will increasingly manifest as a paradox: an ever-improving documentation of an ever-diminishing experience.

If the implications of diminished right-hemispheric attention seem theoretical, evidence from less technologically mediated eras reveals what is at stake. A pre-millennial study on doctor-patient relationships among primary care physicians reveals striking differences in communication styles and their correlation with malpractice claims. Physicians with no claims were found to spend more time with their patients (18.3 minutes compared to 15.0 minutes), use more statements of orientation to educate patients about what to expect, engage in humour, and employ more facilitative techniques, such as soliciting patients' opinions, checking understanding, and encouraging dialogue (Levinson et al., 1997). Notably, the researchers found that "the differences between sued and never-sued physicians are not explained by their quality of care or their chart documentation" (1997, p. 553). The decisive factor is whether patients and families who suffered poor outcomes felt the physician was caring and compassionate. Another study (Ambady et al., 2002, p. 9) reinforces this point, finding that "tone of voice rather than just content of communication may be related to surgical malpractice." Sustained, empathetic attention, humour, compassion, and the primacy of tone over content—these are aspects of communication and traits associated with right-hemispheric attention. Ironically, the very thing the left hemisphere's approach seeks to improve—reducing negligence and avoiding litigation through meticulous documentation—exacerbates the problem. These findings underscore not only the importance of right-hemispheric, democratic values and dispositions but also the self-defeating nature of the left hemisphere's totalitarian approach. In the single-minded pursuit of efficiency, true effectiveness is lost.

**Conclusion**

Physicians, like all experts, are expected to justify their decisions through rules, protocols, and explicit reasoning. This is an institutional demand of the megamachine but it is also a deeply human impulse: when confronted with life-and-death choices, doctors seek structure to make their actions accountable—to colleagues, to patients, and to themselves. Yet this need for procedural clarity conceals a deeper truth. Genuine medical expertise, like all mature skill, is not rule-based but intuitive—an art grounded in embodied perception and years of hard-won experience. As thinkers from Aristotle (2009) to McGilchrist recognise, the highest forms of knowledge resist explicit formulation because they depend on the tacit integration of countless particulars. The left-hemispheric ambition to reduce all such judgment into wholly explicit, representational form is therefore misguided: it loses so much of what it seeks to control that even its apparent success is a Pyrrhic victory. Hence its proper and essential role should remain subordinate to the broader vision of the right hemisphere. No dataset can capture the singular constellation of variables—age, temperament, context, relational nuance—that defines an individual case, nor can statistical likelihoods reproduce the discernment that distinguishes true expertise from mere competence. Medicine, at its highest level, depends on a tacit form of knowing—facilitated by right-hemispheric attention—that eludes the very frameworks designed to formalise it. In a world increasingly determined by the logic of the megamachine, this inarticulable wisdom is easily mistaken for its absence.

To reiterate, the kinds of technical 'solutions' that promise to streamline bureaucracy are themselves expressions of the very left-hemispheric mindset that produced the problem. Even as future versions of AI scribes become more sophisticated—able to register pauses, gestures, or tonal subtleties—they will still translate these phenomena into explicit and representational form. However refined their algorithms, such systems remain instruments of epidianoesis: the externalisation of left-hemispheric attention, whereby the implicit, embodied, and relational are *re*-presented as data. The more proficient they become at this task, the further they extend the logic that defines them. At one level, qualitative dimensions of care— empathy, relationship, attunement—continue to be recast in quantifiable terms. At another, it signals a deeper ontological drift: the gradual reconstitution of medicine itself within the worldview of the megamachine. This is the threshold of metadianoesis, where the process of externalisation becomes reflexive and self-enclosing—attention feeding upon its own projections until representation supplants reality. Without deliberate resistance to this momentum, the conditions of care and the cultivation of expertise grounded in intuitive and experiential understanding risk further subordination to the abstract, self-referential operations of the megamachine.

This, finally, is our predicament: the world remade in the image of a narrow mode of attention. Nowhere is this danger more pressing than in medicine, where the logic of automation is wielded with the aim of reducing the irreducible. When technological systems mediate the physician's every act, intuition atrophies and wisdom is reframed as propositional and ultimately outsourced. The automated transcription and summarisation of words spoken during a consultation is a remarkable technical achievement, yet it remains an act of radical abstraction. The paralinguistic dimensions of embodied communication—tone, cadence, gesture, silence—are not omitted because they are deemed irrelevant; they are never comprehended at all. AI scribes, like all technical artefacts, can only ever operate within a narrow slice of reality, interpreting the physician's insights only on their own truncated terms. Their summaries, filtered through layers of algorithmic reduction, present a record that is precise in detail but impoverished in depth. While abstraction is often necessary, the process of deciding what is relevant—what matters—must remain the prerogative of human attention. Because the kind of attention we bring to bear determines not only what we find but what we fail to see.